\RequirePackage{fix-cm}
\documentclass[smallextended]{svjour3}       
\smartqed  
\usepackage{graphicx}
\begin{document}

\title{Development of a sub-mK Continuous Nuclear Demagnetization Refrigerator
}


\author{David Schmoranzer\and
Rasul Gazizulin\and
S\'{e}bastien Triqueneaux\and
Eddy Collin\and
Andrew Fefferman         
}


\institute{D. Schmoranzer \and R. Gazizulin \and S. Triqueneaux \and E. Collin \and A. Fefferman \at
							Universit\'{e} Grenoble Alpes, CNRS Institut N\'{e}el, 25 rue des Martyrs, BP166, 38042 Grenoble cedex 9, France
              \email{david.schmoranzer@neel.cnrs.fr}           
}

\date{Received: date / Accepted: date}

\maketitle

\begin{abstract}
We present the development of a two-stage PrNi$_5$ continuous demagnetization refrigerator at the Institut Néel/CNRS and numerical simulations of its performance. The thermal model used in the simulations is discussed in detail including the likely sources of heating. We demonstrate the effects of the critical thermal links including superconducting heat switches as well as the heat conductivity of the PrNi$_5$, accounting for the dependence of cooling power on the PrNi$_5$ rod diameter. Our simulations show that if care is taken to minimize the thermal resistance between the nuclear stages, a sample temperature of 1~mK can be maintained under a 20~nW heat load.


\end{abstract}

\section{Introduction}
\label{intro}
In modern low temperature physics, access to ultra-low temperatures below 1~mK has become not only a hallmark of leading laboratories, but also a practical necessity for the investigation of many physical phenomena including superfluidity of $^3$He, quantum states of macroscopic objects or dissipation in amorphous matter~\cite{PickettEnss}. Adiabatic nuclear demagnetization~\cite{Lounasmaa,Pobell} of copper or van Vleck paramagnets, combined with $^3$He - $^4$He dilution refrigeration (DR) has become an indispensable technique for reaching these temperatures, despite the limited amount of experimental time available at ultra-low T on standard demagnetization setups. In practice, the experimental window is often comparable to or less than the spin-lattice relaxation times of the materials of interest, leading to insufficient sample thermalization, e.g., a-Si films show hysteresis between warming and cooling curves in the interval 8-20~mK \cite{DPO}. At the lowest temperatures, thermal relaxation time of order 1~hour is observed. At 1~mK, phonon conduction ($\propto T^3$) is expected to drop by a factor of order 1000, leading to much longer relaxation times.

To overcome the limitation given by the time window, a continuous nuclear demagnetization refrigerator (CNDR) has been proposed~\cite{Shirron,Fukuyama}, utilizing two demagnetization stages connected in series via superconducting heat switches. In this manuscript, we discuss the simulated performance of a similar refrigerator and comment on some of the details of its design. In the following, we present our thermal model of the CNDR setup, details of the numerical calculation and finally discuss the results of the simulations and their implications for the design of the CNDR.
\begin{figure}[tb]
\includegraphics[width=0.99\linewidth]{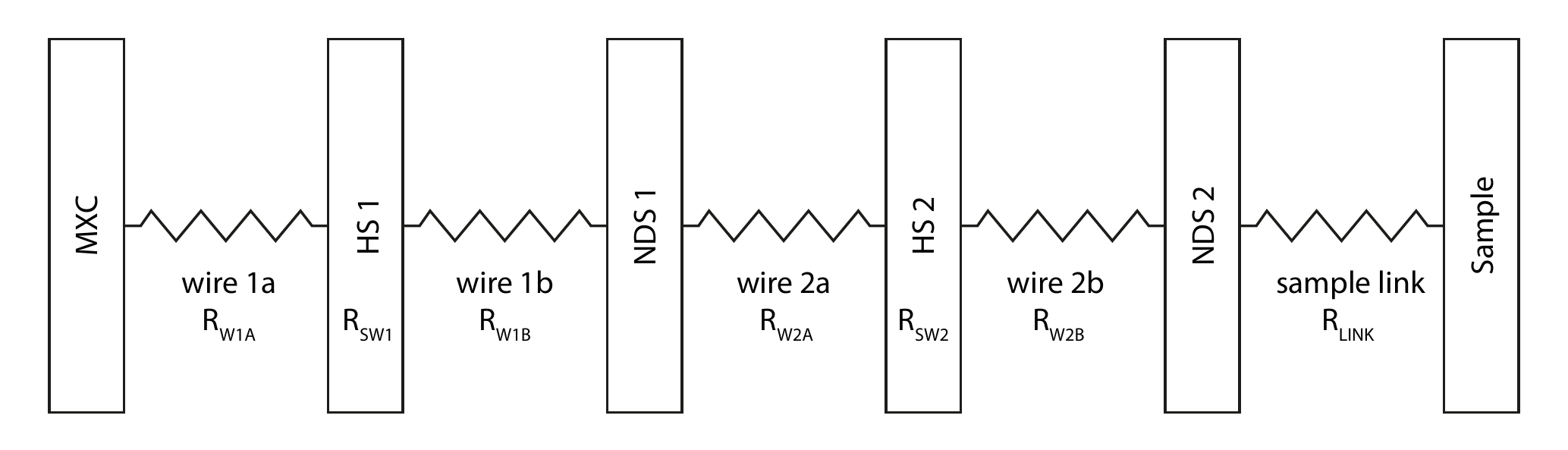}
\caption{Thermal model of the CNDR setup. MXC - mixing chamber of a dilution regfrigerator, HS - heat switch, NDS - nuclear demagnetization stage. Of the resistive wires shown in the setup, only those connecting the nuclear stages to the heat switches were configured to have a significant thermal resistance. Wire 1a and the sample link, represent copper blocks and have negligible thermal resistance in contrast to the thermal links connecting the nuclear stages to heat switches. Additionally, contact resistances were considered at each connection. Where available, these were estimated from the literature~\cite{Mueller,Blondelle}, but for contacts of wires to PrNi$_5$, crude estimates had to be made.}
\label{fig:schematic}       
\end{figure}

\section{Thermal Model}
\label{sec:1}
We consider a model CNDR system with two identical demagnetization stages, each containing 0.2~mol of PrNi$_5$ connected to the mixing chamber of a DR and between themselves via linking elements such as Cu or Ag wires and Al heat switches. The schematics of the system are shown in Fig.~\ref{fig:schematic}.  

In the following, we are assuming that each stage consists of several rods of a fixed diameter, which are all connected to the thermal link leading to the heat switches. For the purposes of the numerical model, this is equivalent to a single rod of said diameter, with its length equal to the sum of the lengths of all constituent rods.

Initially, the two demagnetization stages were thus modeled each as a single entity, characterized only by the respective nuclear and electronic temperatures, $T_n$ and $T_e$, with the heat exchange between the nuclear spins and the electrons given by the Korringa law~\cite{Lounasmaa}. However, a simple estimate of the thermal relaxation time $\tau_{\rm rod}$ of a typical cylindrical rod of PrNi$_5$ shows that this is not likely to be the case in practice. For a rod of radius $R$, the thermal relaxation time is $\tau_{\rm rod} = C \rho R^2 / \kappa$, where $C$ is the heat capacity in constant magnetic field, $\rho$ the density of PrNi$_5$ and $\kappa$ its thermal conductivity~\cite{Folle1981,Meijer}. For a standard 6~mm rod at ultra-low temperatures, this can easily exceed 1~hour due to the low conductivity of the PrNi$_5$ alloy. The thermal gradient inside the nuclear refrigerant will limit the available cooling power of the rods and a more detailed model is necessary to take this into account. 

We therefore use a discrete layered model of each rod (typically with 5, 10, or 20 co-axial cylindrical shells) where each layer is treated separately and has unique nuclear and electronic temperatures. The layers transfer heat to their neighbours as determined by the electronic heat conductivity of PrNi$_5$. Only the outermost layer of the rod is in contact with the thermal link to the heat switches or to the sample volume. Tests showed that 10 layers are sufficient and that the results obtained with 20 do not differ significantly.

The heat flux, $\dot{Q}$, between the various parts of the setup is modeled using the thermal conductance $G(T)$ of the linking element and the temperatures of the end points, $T_1$ and $T_2$ as $\dot{Q} = G(T_{\rm m}) [T_2 - T_1]$, with $T_{\rm m} = (T_1 + T_2) / 2$. While this relation is not strictly true for heat switches in superconducting state, it was nonetheless used for its simplicity, as the error is given by a factor of order unity whereas the leakage in the open state of the switch is several orders of magnitude below other considered heat leaks. The thermal conductance, $G(T)$, of a linking element is defined by the thermal conductivity $\kappa(T)$ of the element material, by the cross-section $A$, length $L$, and by the contact thermal resistances $R_1(T)$, $R_2(T)$ at both ends of the element:
\begin{equation} 
\frac{1}{G(T)} = R_1(T) + \frac{L}{\kappa(T) A} + R_2(T).
\label{eq:conductivity}
\end{equation}
For the Al heat switches, separate relations were employed for their normal and superconducting state, describing heat conduction by electrons ($\kappa \propto T$) and phonons ($\kappa \propto T^3$). The contact thermal resistances $R_1(T)$, $R_2(T)$ were modeled using a temperature-independent electrical resistance and the Wiedermann-Franz law. The cooling power, $\dot{Q}_{\rm c}$, at the mixing chamber plate of the DR is estimated conservatively based on the performance of a Bluefors LD-400 dilution refrigerator:
\begin{equation} 
\dot{Q}_{\rm c} = \alpha T^2 - \dot{q}_0,
\label{eq:DR}
\end{equation}
with the exact values of parameters given by $\alpha = 0.05838$~W K$^{-2}$ for the temperature coefficient and $\dot{q}_0 = 3.854$~$\mu$W for the internal heat leak, resulting in a base temperature of 8.125~mK. We note that a base temperature of 6.5~mK has been reached with the LD-400 installed in CNRS Grenoble. The resistances were chosen so that the total resistance between NDS1 and NDS2 is equivalent to 50, 150, or 500 n$\rm{\Omega}$.

Several types of parasitic heat leaks were included in the model. First, a constant heat leak at the sample volume was specified, with values of 5, 10, or 20~nW. Additionally, for each demagnetization stage in changing magnetic field $B$, three types of heat leaks were considered: (1) eddy current heating, $\dot{Q}_{eddy} = 0.03 \rm{[W T^{-2} s^2]} \dot{B}^2$, (2) vibrational heating, $\dot{Q}_{vibr} = 10^{-8} \rm{[W T^{-1}]} \left|B\right|$, and (3) constant heat leak, $\dot{Q}_{const} = 2$~nW. For eddy current and vibrational heating, the proportionality constants were estimated from the measurements of Parpia et al.~\cite{ParpiaRSI}. We would like to note that vibrational heating is likely to be higher when the CNDR is implemented on a dry dilution unit. Excess heating such as radiation from the magnet, or conduction along centering rings is included in the 2~nW constant heat leak.

\section{Numerical simulations}
\label{sec:2}
The computations were performed by means of integrating the differential equations relating the changes in magnetic fields, changes in temperature, heat fluxes and heat leaks. The code was programmed in Python, with the integration performed using the odeint solver of the SciPy package.

The configuration of the setup was stored in an INI file, and the demagnetization cycles were scripted in a simple custom language providing basic commands for the manipulation of magnetic fields and heat switches. The stability of the computation was governed by the maximum time step permitted for the odeint solver. Introducing the layered model of the demagnetization stages has significantly affected the requirements for the maximum permissible time step to keep the code stable. The initial conditions were chosen so that the entire setup starts at the stable base temperature of the DR unit and both PrNi$_5$ stages are in zero field. Further evolution of the magnetic fields was fully controlled by the scripts.

\begin{figure}[tb]
\includegraphics[width=0.99\linewidth]{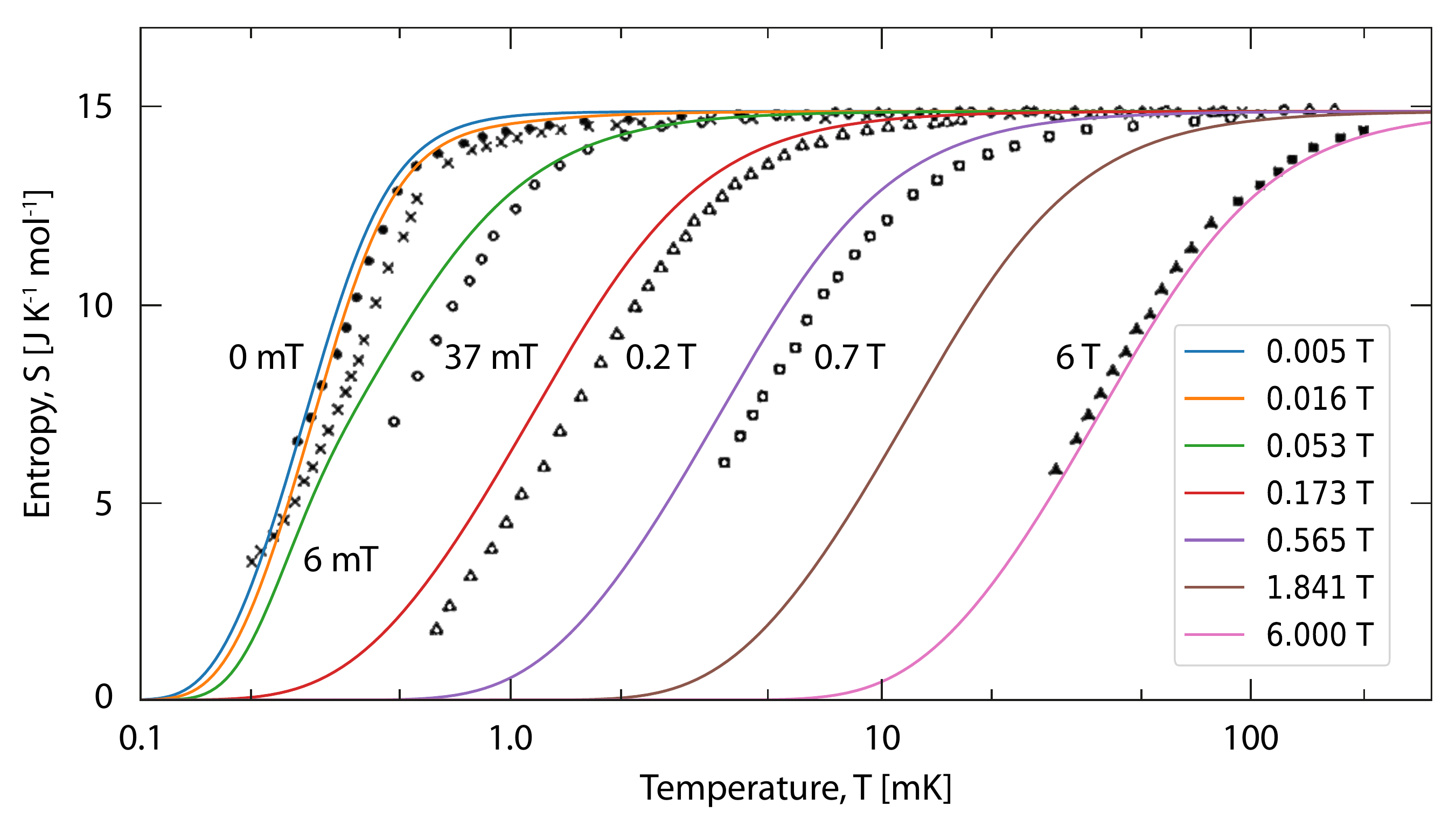}
\caption{Model entropy of PrNi$_5$ (solid lines) superimposed on the measurements presented in Ref.~\cite{KubotaPRL}. Minor discrepancies exist at low fields (c.f. 37~mT), and might affect the final temperatures below 1~mK, but the principal features related to the ordering are captured qualitatively. The model entropy curves are virtually equivalent to paramagnetic ones down to fields $\approx$90~mT. Note that the explicit inclusion of the internal field in the paramagnetic model would shift the zero field curve to 66~mT in contrast to experiment and it cannot yield the increased slope (higher heat capacity) at the transition near 0.4 mK.}
\label{fig:entModel}       
\end{figure}
Initially, the computation used paramagnetic values of entropy for PrNi$_5$, and changes of temperature due to changing magnetic field were calculated using standard expressions~\cite{Lounasmaa}. The following parameters were used for PrNi$_5$: nuclear spin $I = 5/2$, nuclear gyromagnetic ratio $\gamma_n = 1.71$, Knight shift $K = 11.2$, internal field $b = 66$~mT~\cite{Lounasmaa,Pobell,KubotaPRL}, and Korringa constant $\kappa_K = 0.001$ (an upper estimate from Ref.~\cite{Pobell}). Even with the internal field $b$ accounted for, this approach led to difficulties at the lowest temperatures, as the spontaneous ordering of PrNi$_5$ occurring around 0.4~mK is not described within the paramagnetic model. 

Therefore, a mathematical model of the entropy of PrNi$_5$ was constructed based on the measurements of Kubota et al.~\cite{KubotaPRL}, with limits corresponding to the paramagnetic behavior at high T and a steeper dependence of entropy on temperature near the ordering temperature, approximating the experimental data on entropy and specific heat. The resulting lines of constant magnetic field $B$ are shown in the temperature-entropy diagram in Fig.~\ref{fig:entModel} and compared with the experimental data of Ref.~\cite{KubotaPRL}. This model was then used to calculate the nuclear demagnetization properties of the two stages: In each time step, first, the adiabatic change in temperature due to the changing magnetic field was evaluated from the model entropy and then, non-adiabatic effects due to heat exchange and heat leaks were added.

\begin{figure}[tb]
\centering
\includegraphics[width=0.7\linewidth]{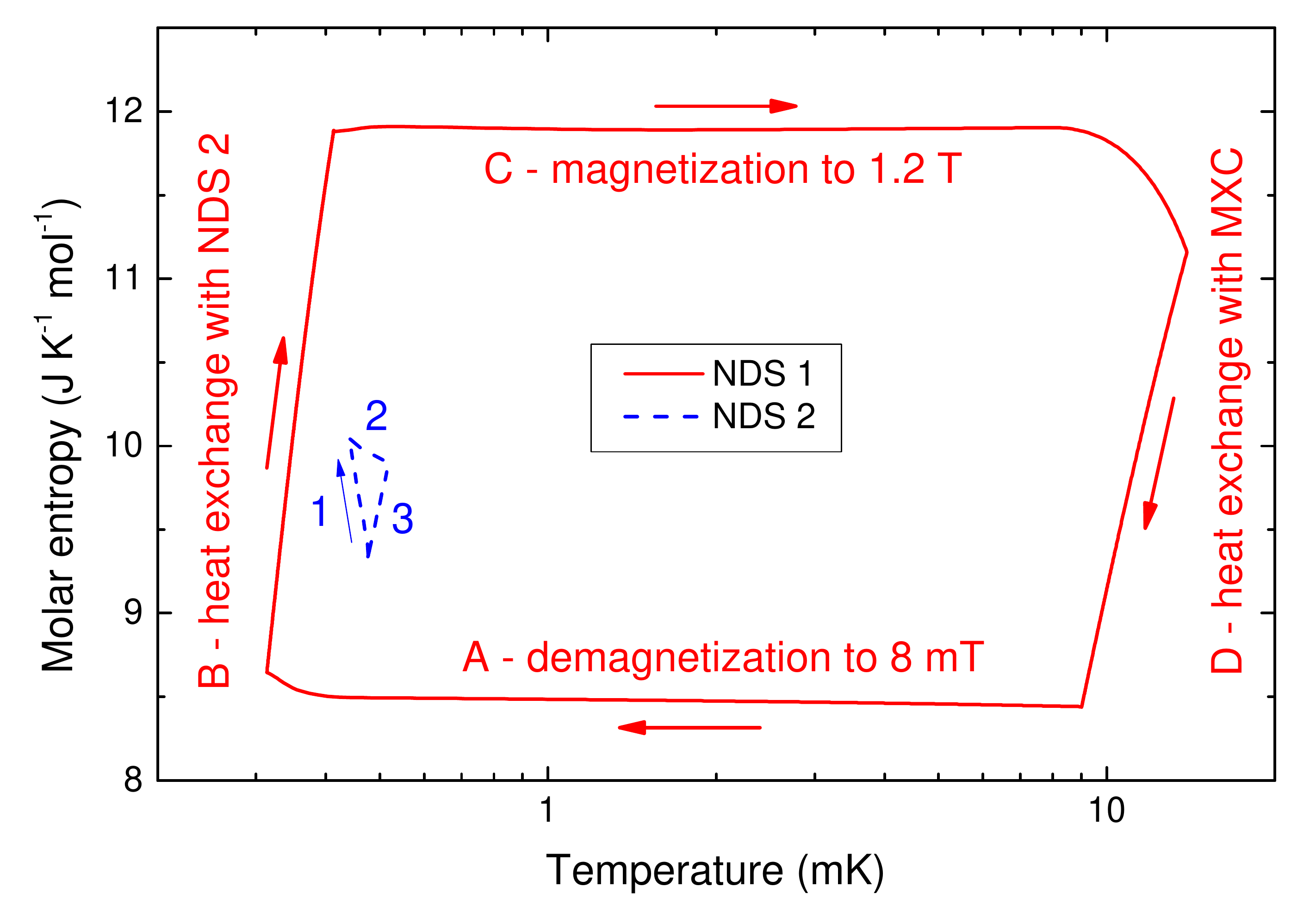}
\caption{Thermal cycles of the first (solid, red) and second (dashed, blue) nuclear demagnetization stages (NDS) of the CNDR. NDS 1 cycle: A - demagnetization to 8~mT while thermally decoupled, B - heat exchange with the second stage, C - magnetization to 1.2~T, D - thermal coupling to the mixing chamber. The steps C and D overlap to save time during the magnetization process - the first stage is coupled to the MXC as soon as its temperature exceeds that of the MXC. NDS 2 cycle: 1 - slow demagnetization from 60 to 50~mT while thermally decoupled, 2 - magnetization to 60~mT while connected to the first stage, 3 - continued cooling by the first stage at constant magnetic field.}
\label{fig:ent}       
\end{figure}

\section{Results and Discussion}
\label{sec:3}
\begin{figure}[tb]
\includegraphics[width=0.99\linewidth]{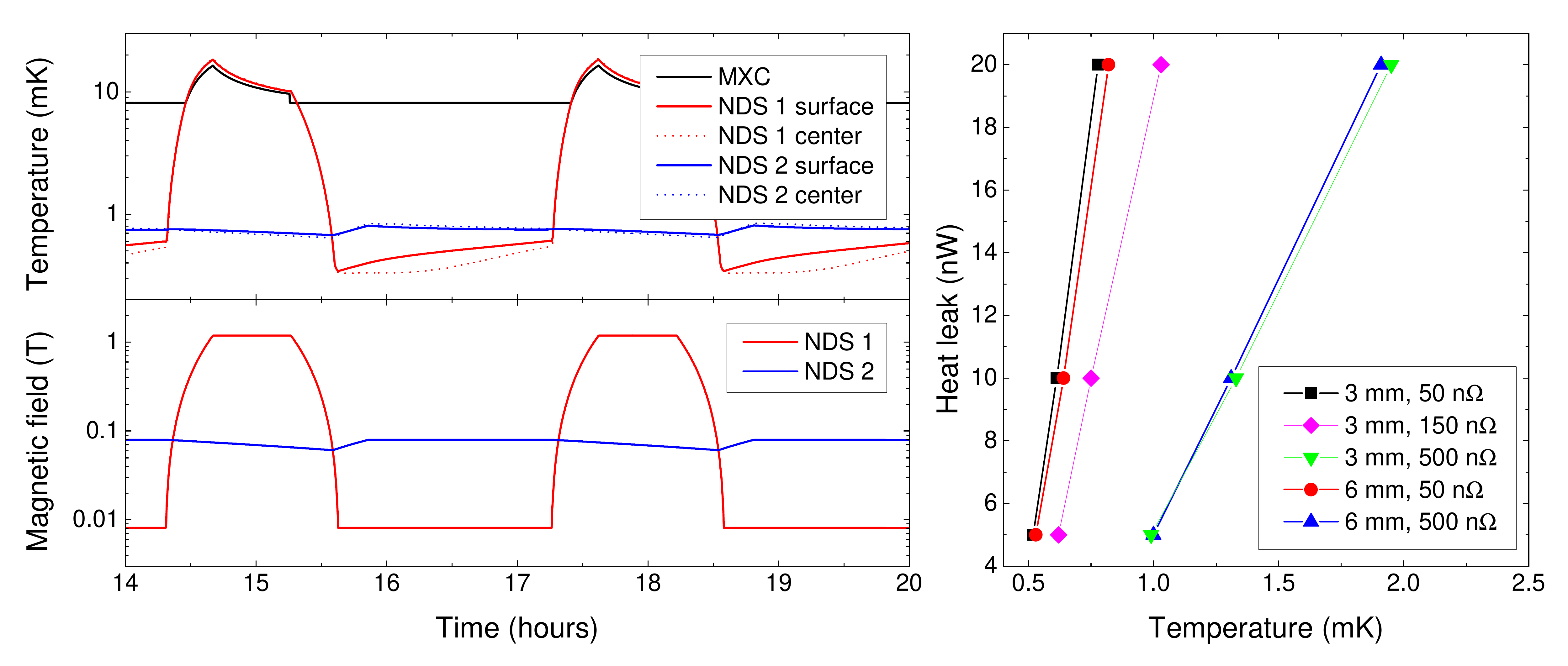}
\caption{Left: Sample time trace of temperatures and magnetic fields of the CNDR setup for 6~mm PrNi$_5$ rods with a 50~n${\rm \Omega}$ resistance of the entire link between NDS 1 and NDS 2, with a heat load of 20~nW on the sample. Right: sample heat load vs. the final temperature attainable with the CNDR setup for the indicated rod sizes and resistances of the link between the stages. The final temperature is the maximum temperature encountered during the cycle of the CNDR. As noted above, the exact final temperatures will likely differ from experiment, due to inaccuracies of the model entropy. Considering the maximum error based on Fig.~\ref{fig:entModel}, the data obtained for 50~n${\rm \Omega}$ resistance still remain below 1~mK.}
\label{fig:timetraces}       
\end{figure}

The thermal cycles of the two stages of the refrigerator are illustrated in the entropy-temperature diagram in Fig.~\ref{fig:ent} for the case of 3~mm rods, 50~n${\rm \Omega}$ resistance and 5~nW heat leak. The durations of the steps were manually adapted for each value of heat leak to produce near-optimum final temperatures. A sample time trace of the relevant temperatures and magnetic fields is shown in Fig.~\ref{fig:timetraces}, together with the dependence of the final temperature on the sample heat load. The final temperature is extracted in each case as the maximum of the second stage temperature within a stable cycle.

The main factor limiting the performance of the CNDR is the slow heat exchange between the two stages when thermally coupled at low T. It is therefore necessary to provide a low resistance path for the heat from one PrNi$_5$ sample to the other. To achieve a value of resistance comparable to 50~n${\rm \Omega}$, optimized contacts of high-conductivity wires with the PrNi$_5$ as well as with the Al heat switch are of particular importance. While suitable solutions for contacting Cu to Al exist \cite{Shigematsu}, our research in this area and practical tests will be reported elsewhere. The more efficient thermalization of the 3~mm rods seems to contribute very little to the performance of the CNDR, as the sample space is only in contact with the outside layer of the rods. Any practical advantage to be gained by replacing the standard 6~mm PrNi$_5$ rods by 3~mm ones is thus related to the increased surface area of the PrNi$_5$ stage and to the correspondingly reduced thermal boundary resistance, which could help us approach the value of 50~n${\rm \Omega}$ for the critical thermal link.

\section{Conclusions}
\label{sec:4}
We have shown that a continuous nuclear demagnetization refrigerator operating with two PrNi$_5$ stages connected in series can reach and maintain temperatures below 1~mK at heat loads up to 20~nW. This makes the operation of such a refrigerator feasible on commercial cryogen-free dilution refrigerators. The key element in the setup is the thermal link between the two stages, including a heat switch and it is crucial to optimize thermal resistances of all its components and their contacts.
The results of our numerical simulations are in general agreement with those of Ref.~\cite{Fukuyama}, but we have used a more detailed description of the demagnetization stages, taking into account their geometry and heat conductivity. In addition, we have used a mathematical model of PrNi$_5$ thermodynamic properties that allow us to describe its refrigeration capabilities in the vicinity of its ferromagnetic ordering occurring near 0.4~mK more precisely than the usual paramagnetic model.

\begin{acknowledgements}
We acknowledge support from the ERC StG grant UNIGLASS No. 714692 and ERC CoG grant ULT-NEMS No. 647917.
\end{acknowledgements}



\end{document}